\documentstyle[prl,twocolumn,aps,epsf]{revtex}
\begin{document}
\title { $Q$-boson interferometry and generalized Wigner function} 
%\date{\today}
%\vspace{-1.35cm}\hfill
%\begin{flushright}
%{\small IFT-P.092/2002}\\
%\end{flushright}
%}
%
\author{Q. H. Zhang$^{a,b}$ and Sandra S. Padula$^c$}
\address{a Physics Department, McGill University,Montreal H3A 2T8,
Canada \\
b Computer Science Department, Concordia University, QCH3G 1M8, Montreal, 
Canada \\
c Instituto de F\'\i sica Te\'orica, Rua Pamplona 145, 01405-900
S\~ao Paulo, Brazil}

\maketitle
\begin{abstract}
Bose-Einstein correlations of
two identically charged $Q$-bosons are derived considering these particles 
to be confined in finite volumes. Boundary effects on single $Q$-boson
spectrum are also studied. We illustrate the effects on the spectrum and 
on the two-$Q$-boson correlation function by means of two toy models. 
We also derive a generalized expression for the 
Wigner function depending on the deformation parameter $Q$, which is 
reduced to its original functional form in the limit of $Q \rightarrow 
1$. 
\end{abstract}
%\vskip -1.35cm

\section{Introduction}

\bigskip
	Two-boson interferometry has for a long time been linked to 
high energy heavy-ion collisions as one of the tools to probe the 
existence of a new phase of matter of
strongly interaction particles, the quark-gluon plasma (QGP), at high
temperature and high baryon density\cite{Wong1,QM}. The hope of 
discovering
the QGP in high energy heavy-ion collisions is to some extent connected
to the possibility of measuring the geometrical sizes of the emission 
region
of secondary particles. And that is the connection point, i.e., 
so-called Hanbury-Brown-Twiss (HBT) interferometry\cite{HBT,GGLP} 
method, 
originally proposed in the 50's for measuring stellar radii. This 
method 
has been largely studied over the last twenty years, and 
has extensively been developed and improved ever since\cite{boal}.

\bigskip

In a previous paper\cite{our1}, we have studied boundary effects on the 
single-particle distribution and on the two-particle correlation 
function, 
motivated by the need to consider more realistic finite systems, and by 
the idea suggested in Ref.\cite{shu}. In that reference, it was shown that 
in heavy-ion collisions the pion system could be thought 
as a liquid of quasi-pions subjected to 
a surface tension. Naturally, it would be expected that 
this surface tension would affect the spectrum distribution, which was 
shown in Ref.\cite{our1,MW93,YS94,MW95,AS97,sarkar}. 
As pion interferometry is sensitive to the geometrical
size of the emission region as well as to the underlying dynamics, we 
would
expect that the boundary would also affect the correlation function, 
which was indeed demonstrated in Ref.\cite{our1}. 

Some time ago, on the other hand, the concept of {\it quons} was 
suggested\cite{greenberg} in 
association to a {\sl  deformation parameter}, $Q$, which was viewed 
as an effective parameter able to encapsulate many essential features 
of complex dynamics of different systems. 
(We call the attention to the notation adopted throughout the paper. 
We use capital $Q$ to 
refer to the bosons under study here for avoiding confusion with the 
relative momentum of the bosonic pairs, $q=p_1 - p_2$, commonly used 
in interferometry and adopted here as well). Effectively, the 
way it works is by reducing the complexity of the interacting systems 
under study into simpler relations, nevertheless at the expense of 
deforming their commutation relations, and thus making 
these more complicated. This is known as $Q$-deformed algebras, 
an approach which has been widely studied 
in statistical physics\cite{qboson1} and also 
in heavy-ion collisions\cite{agp}.  Particularly interesting 
is the approach in Ref.\cite{gastao}, where it was shown that 
the composite nature of the particles (pseudo-scalar mesons) under 
study could result into $Q$-deformed structures linked to the deformation 
parameter $Q$. In that reference this parameter is then interpreted 
as a measure of effects coming from the internal degrees of 
freedom of composite particles (mesons, in our case), being 
the value of $Q$ dependent on the {\sl degree of overlap} 
of the extended structure of the particles in the medium. Being so, 
the $Q$-parameter could be related to the power of {\sl probing 
lenses}, for mimicking the effects of internal constituents of the 
bosons. In this case, and for high enough {\sl magnification}, 
the bosonic behavior of the $Q$-bosons could be blurred 
by the fermionic effect of their internal constituents, 
which would result in decreasing the value of $Q$. 
We will see that our results are also compatible with 
this interpretation.

    In view of our previous study of confined pions subjected 
to finite size boundaries, and of the $Q$-boson approach mentioned 
above, we realized it would be very interesting to analyze its  
effects on the two-identical Q-boson correlation function. Besides, 
adding this extra degree of freedom extends and generalizes our 
previous approach. Along these lines, 
Ref.\cite{agp,AGI} turned out to be of special interest 
to our investigation.  However, in those references the approach 
was focused on the intercept, $\lambda$, of the 
two-particle correlation function at zero momentum 
difference, (i.e., $\lambda = C(q=0,K)-1)$, and restricted to 
single modes only. All the possible consequences on 
the effective geometrical information, which are generally even more 
interesting, were completely neglected. In this paper we  
develop full $Q$-boson two-particle interferometric relations and 
simultaneously study the additional effects caused by the finite size 
boundary on the $Q$-boson spectrum and on the two-$Q$-boson correlation function.

\bigskip
The plan of this paper is as follows:  in section II, we
derive the $Q$-boson single-inclusive distribution, as well as the 
two-$Q$-boson
correlation function, considering a density matrix suited for
describing charged identical $Q$-boson correlation effects.
In section III, the boundary effects on the two-$Q$-boson correlation 
and single
particle spectrum distribution are illustrated by means of two simple 
specific examples. The conclusions are discussed in section IV. 
Finally, we discuss two complementary topics in Appendices I and II. 
In the first one, we discuss the limit $Q \rightarrow 0$ in detail. 
In Appendix II, we re-derive the relations for the single-  and 
two-particle distributions, as well as the generalized version 
of the Wigner function for another type of $Q-$boson\cite{agp,AGI},  
different than the one discussed in the body of the paper.

\section{Spectrum and Two-$Q$-boson correlation function}

In this section, we derive general relations for describing 
the single- as well as the two-particle inclusive distributions, 
which
would be suited for describing charged $Q$-bosons bounded in a finite 
volume. 
For doing this, we extend the hypotheses assumed in Ref.\cite{our1} to 
the pions considered here as a $Q$-boson system. Essentially, these 
could be 
summarized as follows: 
the effects of interaction among the $Q$-bosons could be modeled by 
considering 
that they move in an attractive mean field potential, which extends 
over the whole system. In the two-(quasi)particle case, this implies 
that they would not suffer any other effects besides 
the mean field attraction and the identical particle symmetrization. 
The effect due to the fermionic (constituents) internal degrees of 
freedom, 
along the lines suggested in Ref.\cite{gastao}, if any, would be 
represented 
by the effective deformation parameter Q. In the present analysis, 
as assumed before in Ref. \cite{our1}, the pions represented 
by $Q$-bosons are considered to be quasi-bound in the system, with the 
surface tension\cite{shu} acting as a reflecting boundary. 
The $Q$-boson wave function could then be considered as 
vanishing outside this boundary. Once more, we assume that 
these particles become free when their average separation is larger 
than their interaction range and we consider this 
transition to happen very rapidly, in such a way that the momentum 
distribution of the $Q$-bosons could be essentially governed by their 
momentum distribution just before they freeze out. We then study the 
modifications on the observed $Q$-boson momentum distribution caused 
by the presence of this boundary. On the other hand, we know that 
interferometry is sensitive to the geometrical size of the emission 
region 
as well as to the underlying dynamics, and, we would 
expect that the boundary would also affect the correlation function, 
similarly as it affected the pions in \cite{our1}. However, as we shall 
see later, there is a significant difference in the present case: the 
parameter $\lambda$, i.e., the intercept (at ${\bf q=0}$) of the 
two-particle correlation function,  
will be considerably different as compared to 
the case of a normal (i.e., in the 
limit of $Q \rightarrow 1$) pion, but will be recovered in the 
appropriate limit. 

For deriving the relations that allow to describe the single- and 
two-particle
inclusive distributions, we start by assuming that the $Q$-boson 
creation 
operator in coordinate space can be expressed by \cite{our1}
\begin{eqnarray}
\hat{\psi}^{\dagger}({\bf x})=
\sum_{l}\hat{a}^{\dagger}_{l}\psi^*_{l}({\bf x}),
\end{eqnarray}
where $a^{\dagger}_{l}$ is the creation operator for creating a
$Q$-boson in a quantum state characterized by a
quantum number $l$. Then,
$\psi_{l}({\bf x})$ is one of eigenfunctions belonging to
a localized complete set, which
satisfies the orthonormality condition
\begin{equation}
\int d{\bf x} \psi^*_{l}({\bf x})
\psi_{l'}({\bf x})=\delta_{l,l'},
\end{equation}
and completeness relation
\begin{equation}
\sum_{l}\psi^*_{l}({\bf x})\psi_{l}({\bf y})=
\delta({\bf x}-{\bf y}).
\label{xx3}
\end{equation}

Similarly, the $Q$-boson annihilation operator in coordinate space can 
be
written as
\begin{equation}
\hat{\psi}({\bf x})=\sum_{l}\hat{a}_{l}\psi_{l}({\bf x}).
\end{equation}

In momentum space, the corresponding $Q$-boson creation
operator, $\hat{\psi}^{\dagger}({\bf p})$, and
annihilation operator, $\hat{\psi}({\bf p})$,
can be expressed, respectively, as
\begin{equation}
\hat{\psi}^{\dagger}({\bf p})=\sum_{l} \hat{a}^{\dagger}_{l}
\tilde{\psi}^*_{l}({\bf p})
\end{equation}
and
\begin{equation}
\hat{\psi}({\bf p})=\sum_{l} \hat{a}_{l}
\tilde{\psi}_{l}({\bf p}) ,
\end{equation}
where
\begin{equation}
\tilde{\psi}_{l}({\bf p})=\frac{1}{(2\pi)^{3/2}}
\int \psi_{l}({\bf x})e^{i{\bf p}\cdot {\bf x}}d{\bf x}   .
\label{Fourier}\end{equation}

The $Q$-bosons are then defined by means of the algebra 
satisfied by their creation and annihilation operators, 
i.e.,  \cite{AGI}

\begin{eqnarray}
&&a_{l}a_{l'}^{\dagger}-Q^{\delta_{l,l'}}
a_{l'}^{\dagger}a_{l}=\delta_{l,l'}
\nonumber\\
&&[a_{l},a_{l'}]=[a_{l}^{\dagger},a_{l'}^{\dagger}]=0,
\nonumber\\
&&[\hat{N}_{l},a_{l'}]=-\delta_{l,l'}a_{l}
\nonumber\\
&&[\hat{N}_{l},a_{l'}^{\dagger}]=
\delta_{l,l'}a_{l}^{\dagger},
\nonumber\\
&&[\hat{N}_{l},\hat{N}_{l'}]=0,
\nonumber\\
\label{commutators}\end{eqnarray}
Here $Q$ is a (C-number) parameter, assumed to be within 
the interval $[-1,1]$, and $\hat{N}_{l}$ is the number operator, 
which can be expressed as
\begin{equation}
\hat{N}_{l}=\sum_{s=1}^{\infty}\frac{(1-Q)^s}{(1-Q^s)}
(a_{l}^{\dagger})^s (a_{l})^{s}. 
\label{nlambda}\end{equation}
It can be easily verified that, for $Q=1$, the normal bosonic limit  
is recovered, i.e., the particles then 
obey the regular bosonic commutation relations, as it would be 
expected.

We write the density matrix operator for our $Q$-bosonic system
as

\begin{eqnarray}
\hat{\rho}&=&\exp\left[-\frac{1}{T} (\hat{H}-\mu \hat{N})\right]
=\prod_{l} \rho_{l},
\nonumber\\
\rho_{l}&=&\exp\left[-\frac{1}{T} (\hat{H}_{l}-\mu \hat{N}_{l})\right]
\; \;, \label{rho}\end{eqnarray}
where
\begin{equation}
\hat{H}=\sum_{l}\hat{H}_{l}; \; \; \; \;
\hat{H}_{l}=E_{l}\hat{N}_{l}; \; \; \; \;
\hat{N}=\sum_{l}\hat{N}_{l}
\; \;, \label{HN}\end{equation}
are the Hamiltonian and number operators, respectively; T is the 
temperature.

 	The corresponding normalization is explicitly included
in the definition of the expectation value of observables as,
for instance, for an operator $\hat{A}$

\begin{equation}
\langle \hat{A}\rangle =\frac{tr\{ \hat{\rho} 
\hat{A}\}}{tr\{\hat{\rho}\}}
\; \;.\label{A}\end{equation}

With the above definitions, it is easy to verify that
\begin{equation}
%%tr(\rho_{l})=\frac{1}{1-\exp[-\frac{1}{T}(E_l-\mu_{l})]}
%tr(\rho_{l})= \sum_{n} {_{_\l}\!\langle} n | \rho_\l | n \rangle_{_\l} = 
%\frac{1}{1-\exp[-\frac{1}{T}(E_l-\mu)]}
tr(\rho_{l})= \sum_{n} {_\l}\!{\langle} n | \rho_{\l} | n \rangle_{\l} = 
\frac{1}{1-\exp[-\frac{1}{T}(E_l-\mu)]}
\; ,\end{equation}
where 
\begin{equation}
| n \rangle_{\l} = \frac{1}{\sqrt{[n]!}} (a_{\l}^{\dagger})^n |0 
\rangle \; \; ; 
\; \; [n] = \frac{1-Q^n}{1-Q}
.\label{nstate}\end{equation}

From the above equations, we can compute the expectation values 
\begin{equation}
\langle a_{l}^{\dagger}a_{l}\rangle
=
\frac{1}{ \exp[\frac{1}{T}(E_{l}-\mu)]-Q}
\; ,\label{a+a}\end{equation}
and
\begin{equation}
\langle a_{l}^{\dagger}
a_{l}^{\dagger}
a_{l}
a_{l}\rangle
=
\frac{1+Q}{\{e^{\frac{1}{T}(E_{l}-\mu)}-Q\}
\{e^{\frac{1}{T}(E_{l}-\mu)}-Q^2 \}}
\;.\label{a+a+aa}\end{equation}

 Then, the single $Q$-boson distribution can be
written as
\begin{eqnarray}
P_1({\bf p})&=&\langle \hat{\psi}^{\dagger}({\bf p})\hat{\psi}({\bf 
p})\rangle
\nonumber\\
&=&\sum_{l}\sum_{l'}
\tilde{\psi}_{l}^*{(\bf p)}
\tilde{\psi}_{l'}{(\bf p)}
\langle \hat{a}^{\dagger}_{l}\hat{a}_{l'}\rangle
\; \;.\label{P1in}\end{eqnarray}

The expectation value
$ \langle \hat{a}^{\dagger}_{l}\hat{a}_{l'} \rangle $
is related to the occupation probability of a 
single-particle state $l$, 
$N_l$,  by the following relation 
\begin{equation}
\langle \hat{a}^{\dagger}_{l}\hat{a}_{l'} \rangle
=\delta_{l,l'}N_{l}
\; \; .\label{adagaa}\end{equation}
For a $Q$-bosonic
system in equilibrium at a temperature $T$ and chemical
potential $\mu$, $N_l$
is represented by the modified Bose-Einstein distribution
\begin{equation}
N_{l}=\frac{1}{\exp{\left[\frac{1}{T} (E_{l}-\mu)\right]}-Q }
\; \;.\label{Nlamb}\end{equation}

By inserting Eq. (\ref{adagaa})  and (\ref{Nlamb}) into (\ref{P1in}), 
we
obtain the single-particle spectrum for one $Q$-boson species as
\begin{equation}
P_1({\bf p})=\sum_{l}N_{l}
\tilde{\psi}^*_{l}({\bf p})
\tilde{\psi}_{l}({\bf p})
\;.\label{p1g}\end{equation}

Similarly, the two-$Q$-boson distribution function can be written as
\begin{eqnarray}
P_2({\bf p_1,p_2})&=&
\langle  \hat{\psi}^{\dagger}({\bf p_1})\hat{\psi}^{\dagger}({\bf p_2})
\hat{\psi}({\bf p_1})\hat{\psi}({\bf p_2})\rangle
\nonumber\\
&=&\sum_{l_1,l_2,l_3,l_4}
 \tilde{\psi}_{l_1}^*({\bf p_1})
\tilde{\psi}_{l_2}^*({\bf p_2})
\tilde{\psi}_{l_3}({\bf p_1})
\tilde{\psi}_{l_4}({\bf p_2})
\nonumber\\
&&
\langle  \hat{a}^{\dagger}_{l_1}\hat{a}^{\dagger}_{l_2}
\hat{a}_{l_3}\hat{a}_{l_4} \rangle
\nonumber\\
&=&\sum_{l_1,l_2,l_3,l_4}
\tilde{\psi}_{l_1}^*({\bf p_1})
\tilde{\psi}_{l_2}^*({\bf p_2})
\tilde{\psi}_{l_3}({\bf p_1})
\tilde{\psi}_{l_4}({\bf p_2})
\nonumber\\
&&
\left[\langle \hat{a}^{\dagger}_{l_1}\hat{a}_{l_3}\rangle
\langle \hat{a}^{\dagger}_{l_2}\hat{a}_{l_4}\rangle_{l_1 \ne l_2}
\right.
\nonumber\\
&&
+ \langle \hat{a}^{\dagger}_{l_1}\hat{a}_{l_4}\rangle
\langle \hat{a}^{\dagger}_{l_2}\hat{a}_{l_3}\rangle_{l_1 \ne l_2}
\nonumber\\
&&
\left. +\langle \hat{a}^{\dagger}_{l_1}\hat{a}^{\dagger}_{l_2}
\hat{a}_{l_3}\hat{a}_{l_4}\rangle_{l_1=l_2=l_3=l_4}
\right ]
\nonumber\\
&=&P_1({\bf p_1})P_1({\bf p_2})+
|\sum_{l}N_{l}\tilde{\psi}_{l}^*({\bf p_1})
\tilde{\psi}_{l}({\bf p_2})|^2
\nonumber\\
&&
+\sum_{l}
\tilde{\psi}_{l}^*({\bf p_1})
\tilde{\psi}_{l}^*({\bf p_2})
\tilde{\psi}_{l}({\bf p_1})
\tilde{\psi}_{l}({\bf p_2}) \times
\nonumber\\
&&
[\langle \hat{a}^{\dagger}_{l}\hat{a}^{\dagger}_{l}
\hat{a}_{l}\hat{a}_{l}\rangle 
-2\langle \hat{a}^{\dagger}_{l}\hat{a}_{l}\rangle^2]
\label{p2g}
\end{eqnarray}

Using Eq.(\ref{a+a}) and (\ref{a+a+aa}), we finally have
\begin{eqnarray}
P_2({\bf p_1,p_2})&=&
P_1({\bf p_1})P_1({\bf p_2})+
|\sum_{l}N_{l}\tilde{\psi}_{l}^*({\bf p_1})
\tilde{\psi}_{l}({\bf p_2})|^2
\nonumber\\
&&
- \sum_{l}
\tilde{\psi}_{l}^*({\bf p_1})
\tilde{\psi}_{l}^*({\bf p_2})
\tilde{\psi}_{l}({\bf p_1})
\tilde{\psi}_{l}({\bf p_2})\times
\nonumber\\
&&
(1-Q)\cdot N_{l}^2
\frac{\exp[\frac{1}{T}(E_{l}-\mu)]+Q}
{\exp[\frac{1}{T}(E_{l}-\mu)]-Q^2}
\label{p2ff}
\end{eqnarray}

The two-particle correlation can then be written as
\begin{eqnarray}
&&C_2({\bf p_1,p_2})=\frac{P_2({\bf p_1,p_2})}{P_1({\bf p_1})
P_1({\bf p_2})} = 
\nonumber\\
&&=1+
\left\{ \sum_{l}N_{l}|\tilde{\psi}_{l}({\bf p_1})|^2
\sum_{l}N_{l}|\tilde{\psi}_{l}({\bf p_2})|^2 \right\}^{-1} 
\times 
\nonumber\\
&&
\; \; \; \sum_{l,l'} \; \ N_{l} N_{l'} 
\tilde{\psi}_{l}^*({\bf p_1}) \tilde{\psi}_{l'}^*({\bf p_2}) 
\tilde{\psi}_{l}({\bf p_2}) \tilde{\psi}_{l'}({\bf p_1})\times
\nonumber\\
&&
\left\{ 1 - \delta_{l,l'} (1-Q)\cdot  
\frac{\exp(\frac{1}{T}(E_{l}-\mu))+Q}
{\exp(\frac{1}{T}(E_{l}-\mu))-Q^2} \right\} 
\; \;. \nonumber\\
\label{c21st}
\end{eqnarray}

It is interesting to note that, for Q=1, we regain the results
in Ref\cite{our1}. Moreover, for Q=0, we also get identical 
results as shown in the Appendix A of that reference, 
corresponding to {\sl classical} Boltzmann distribution, 
for either single or multi-modes. 
Nevertheless, the naively expected classical limit of 
$C_2({\bf p_1,p_2}) \rightarrow 1$ is recovered 
for single mode only, independently on the values of 
${\bf p_1}$ and ${\bf p_2}$.   

In analogy to the common practice in bosonic interferometry, it is natural 
to introduce  the intercept parameter, $\lambda$,
by means of the relation  $\lambda = C({\bf q=0},{\bf K})-1)$
i.e., as the intercept of the two-particle correlation function from which 
the classical limit is subtracted 
(this procedure, however, has {\sl no relation} to the historical interpretation 
of $\lambda$ as a parameter signaling either total chaoticity or 
partial coherence of the 
emitting source). 

From Eq. (\ref{c21st}), it is straightforward to show that, 
when ${\bf q}={\bf p_1}-{\bf p_2}=0$, and, consequently, 
${\bf K}=\frac{1}{2}({\bf p_1}+{\bf p_2})={\bf p_1}={\bf p_2}$
\begin{eqnarray}
&&C_2({\bf K,K})=2 - 
\frac{1}{\sum_{l}N_{l}|\tilde{\psi}_{l}({\bf K})|^2
\sum_{l}N_{l}|\tilde{\psi}_{l}({\bf K})|^2}\times
\nonumber\\
&&
[\; (1-Q)\sum_{l}
\tilde{\psi}_{l}^*({\bf K})
\tilde{\psi}_{l}^*({\bf K})
\tilde{\psi}_{l}({\bf K})
\tilde{\psi}_{l}({\bf K})
\cdot N_{l}^2
\nonumber\\
&&
\times
\frac{\exp(\frac{1}{T}(E_{l}-\mu))+Q}
{\exp(\frac{1}{T}(E_{l}-\mu))-Q^2} \; ]
\label{c22nd}
\;. \end{eqnarray}

We see that, for $Q=1$, we regain the ideal result for the bosonic 
intercept at the zero momentum difference, ${\bf q}={\bf p_1}-{\bf 
p_2}=0$, 
i.e., $C_2({\bf K,K})=2$. 
On the other hand, for $Q=0$ we again recover the naively 
expected Boltzmann result, $C_2({\bf K,K})=1$ for single modes only, 
and in particular for ${\bf p_1}={\bf p_2}=K$, as discussed above. 
For multi-modes, however, 
it is not recovered, since there is some remnant communication among 
$Q$-bosons approaching the classical limit in this case, also 
verified in the result presented in the appendix mentioned above. 

From Eq.(\ref{c22nd}) the intercept of the correlation function, 
$\lambda$, can be immediately identified as 
  
\begin{eqnarray}
&&\lambda({\bf K})=C_2({\bf K},{\bf K})-1=1 -
\nonumber\\
&&
\frac{(1-Q)}{\sum_{l}N_{l}|\tilde{\psi}_{l}({\bf K})|^2
\sum_{l}N_{l}|\tilde{\psi}_{l}({\bf K})|^2}
\times
[ \; \sum_{l}
\tilde{\psi}_{l}^*({\bf K})
\tilde{\psi}_{l}^*({\bf K})
\nonumber\\
&&
\times
\tilde{\psi}_{l}({\bf K})
\tilde{\psi}_{l}({\bf K})
\cdot N_{l}^2
\left( \frac{\exp(\frac{1}{T}(E_{l}-\mu))+Q}
{\exp(\frac{1}{T}(E_{l}-\mu))-Q^2} \right) \; ]
\; . \label{chaot} 
\end{eqnarray}
We see that the intercept decreases with decreasing $Q$, being always 
smaller than unity for $0 \le Q \le 1$. 
On the other hand, 
it is interesting to point that the definition for 
the intercept 
parameter given by Eq. (\ref{chaot}) differs from 
the one in Ref.\cite{agp,AGI}, mainly, but not only, because it is 
there 
defined exclusively for a single mode. In that reference, comparison is
made with experimental points for $\lambda$, which has always been in 
the 
limit $0 \le \lambda \le 1$. We could proceed similarly within our 
model as well, by comparing Eq. (\ref{chaot}) to the experimental 
points. 
Nevertheless, we prefer not to do so 
because it is well-known that other factors, such as resonances, 
dynamical and multi-particle effects, as well as kinematical cuts, 
could 
also cause the intercept to drop into that interval.

\vskip.5cm

The above derivation can also be reformulated within the 
Wigner function approach. For doing this, we develop the product of 
four wave-functions $\psi^{(*)}$ in Eq. (\ref{c21st}) into
the product of the corresponding Fourier transforms, leading to

\begin{eqnarray}
&& \tilde{\psi}_{l}^*({\bf p_1})
\tilde{\psi}_{l'}^*({\bf p_2}) \tilde{\psi}_{l}({\bf p_2})
\tilde{\psi}_{l'}({\bf p_1}) = 
\nonumber\\
&&
= 
\int \frac{d^3 r_1}{(2 \pi)^{3/2}}e^{-i{\bf p_1}.{\bf r_1}} 
\psi_{l}^*({\bf r_1})\int \frac{d^3 r_2}{(2 \pi)^{3/2}}
e^{-i{\bf p_2}.{\bf r_2}}\psi_{l}^*({\bf r_2})\times 
\nonumber\\
&&
\int \frac{d^3 r'_2}{(2 \pi)^{3/2}}e^{i{\bf p_2}.{\bf r'_2}} 
\psi_{l'}({\bf r'_2})\int\frac{d^3 r'_1}{(2 \pi)^{3/2}}
e^{i{\bf p_1}.{\bf r'_1}} \psi_{l'}({\bf r'_1}) 
\nonumber\\
&&
=\!\int d^3 x \; e^{-i{\bf q}. {\bf x}}
\int \frac{d^3 \Delta x }{(2 \pi)^3}\; e^{-i{\bf K}.{\bf \Delta x}}
\psi_{l}^*({\bf x}\!+\!\frac{\bf \Delta x}{2})
\psi_{l}({\bf x}\!-\!\frac{\bf \Delta x}{2}) 
\nonumber\\
&&
\times \int d^3 y \; e^{i{\bf q}. {\bf y}}
\int \frac{d^3 \Delta y}{(2 \pi)^3} \; 
e^{i{\bf K}.\Delta {\bf y}}
\psi_{l'}^*({\bf y}\!-\!\frac{\bf \Delta y}{2})
\psi_{l'}({\bf R}\!+\!\frac{\bf \Delta y}{2}) 
, \nonumber\\
\label{N4psi}
\end{eqnarray}
where we have defined ${\bf K=(p_1+p_2)}/2$ as
the two-$Q$-boson average momentum, and ${\bf q=p_1-p_2}$  as their 
relative momentum. For writing the last equality, we have also 
changed variables as follows: ${\bf r_1} - {\bf r_2} = \Delta {\bf x} 
\; ; \;
{\bf r_1} + {\bf r_2} = 2 {\bf x} \; ; \;
{\bf r'_1} - {\bf r'_2} = \Delta {\bf y} \; ; \; {\bf r'_1} + {\bf 
r'_2} = 
2 {\bf y}$.

Then we can define the Wigner function
associated to the state $l$ as 

\begin{equation}
g_{l} ({\bf x},{\bf K}) = 
\int \frac{d^3 \Delta x}{(2 \pi)^3} \; 
e^{-i{\bf K}.{\bf \Delta x}}
\psi_{l}^*({\bf x} + \frac{\bf \Delta x}{2})
\psi_{l}({\bf x} - \frac{\bf \Delta x}{2}) 
\; . \end{equation}

We can proceed analogously to define the equivalent function for the 
integration in ${\bf y}$ and  ${\bf \Delta y}$, remembering that 
$g_{l} ({\bf x},{\bf K}) = g_{l}^*({\bf x},{\bf K})$. Then, 
denoting by 
$$g({\bf x},{\bf K}) = \sum_{l} N_l \; g_l, $$ 
we can finally define the generalized Wigner function of the problem as 

\begin{eqnarray}
&&g ({\bf x},{\bf K} ; {\bf y},{\bf K}) = g({\bf x},{\bf K}) g({\bf 
y},{\bf K}) 
\; - \; (1-Q)\nonumber\\
&&
\sum_{l}\left\{ N_{l}^2\left[\frac{\exp(\frac{1}{T}(E_{l}-\mu))+Q}
{\exp(\frac{1}{T}(E_{l}-\mu))-Q^2}\right] g_{l} ({\bf x},{\bf K}) \; 
g_{l} ({\bf y},{\bf K})\right\}.\nonumber\\
\label{wigner} 
\end{eqnarray}
We see that, for $Q=1$, the above expression is reduced to the usual 
result of the original Wigner function, i.e., $g ({\bf x},{\bf K} ; 
{\bf y},{\bf K}) = g({\bf x},{\bf K}) g({\bf y},{\bf K})$. On the other 
hand, for $Q=0$, Eq. (\ref{wigner}) is identically zero for single 
modes 
only, as it would be expect in the limit of Boltzmann statistics. 
Nevertheless, in the multi-mode case, as already shown in Eq. 
(\ref{c22nd}), 
there seems to be some sort of residual correlation among $Q$-bosons 
even in the classical limit. Aiming at better understanding this limit 
we further explore the $Q \rightarrow 0$ case in the Appendix I.

By means of this Wigner function, the two-$Q$-boson correlation 
function 
can be rewritten as

\begin{equation}
C_2({\bf p_1,p_2})= 1 +
\frac{\int \int e^{-i{\bf q}. ({\bf x}-{\bf y})}
g({\bf x},{\bf K} ; {\bf y},{\bf K}) d{\bf x}d{\bf y}}
{\int g({\bf x},{\bf p_1})d{\bf x} \int g({\bf y},{\bf p_2}) d{\bf y}}
%\nonumber\\
\label{c23rd}\end{equation}

The above generalized Wigner function, $g({\bf x,K})$, can be
interpreted as the probability of finding a
$Q$-boson at a point {\bf x} with momentum {\bf K}.
Differently from previous formulations\cite{pratt,pgg,CGZ} we see that,  
if pions are treated as 
$Q$-bosons under certain regimes, 
there is now an additional term in Eq. (\ref{wigner}). The 
modified two-particle Wigner function no longer can be reduced to the 
Fourier transform of the product of two single-particle Wigner 
functions, 
but acquires an extra term depending on $Q$ in a non-trivial way. 
As a consequence, 
for $0 \le Q \le 1$, we can anticipate that the correlation function 
would be narrower and the intercept parameter, 
$\lambda$, would drop below unity. 
We will illustrate more clearly the effects of the {\sl deformation 
parameter}, 
$Q$, on the correlation function and on the intercept parameter in 
the  next section, by means of two toy models. 
 
In summary, we could say that, maybe under certain circumstances, pions 
produced in heavy-ion collisions could be treated as free particles. 
Nevertheless, and as motivated in the beginning of the present section, 
in many others, the interactions of pions among themselves and with 
other 
particles produced in relativistic heavy-ion collisions may not be 
negligible. In these cases, similar to what has been suggested in 
Ref.\cite{AGI,agp}, what it is proposed here is to mimic those 
interactions by considering pions as $Q$-bosons. 
In particular, the interpretation of $Q$ as an effective parameter 
reflecting the fermionic constituents\cite{gastao} of the $Q$-bosons 
is appealing. Mainly if we consider that unconfined degrees of freedom 
could be produced in high energy heavy ion collisions and manifest 
themselves as $Q$-bosons in the pre-bosonic stages
In this sense, 
they would be regarded as {\it memory traces} from those pre-confined 
stages, just before the boson emission.

\section{Two-$Q$-boson correlation from a finite volume}

\subsection{Toy model }

To explore the effects of the deformation parameter $Q$ and of 
the boundary on the single- and two-$Q$-boson distribution functions, 
we assume that they are confined in a one-dimension box, $[0,L]$,
for simplicity. The three-dimensional extension should be  
straightforward. It can be easily checked that
the corresponding wave-function in the 1-D case is given by 
\begin{equation}
\psi_{k}(x)=\sqrt{\frac{2}{L}}\sin k\cdot x,
\end{equation}
with
\begin{equation}
k\cdot L=n\pi, n=1,2,3...  ~~~~~ .
\end{equation}

Then, the corresponding Fourier transform, 
$\tilde{\psi}_{k}(p)$, can be expressed as
\begin{eqnarray}
\tilde{\psi}_{k}(p)&=&\frac{1}{(2\pi)^{1/2}}
\frac{1}{\sqrt{2L}}[\frac{\exp(i(k-p)L)-1}{p-k}-
\nonumber\\
&&
\frac{\exp(-i(k+p)L)-1}{p+k}], 
\end{eqnarray}
or, equivalently, its square modulus would be written as 
\begin{eqnarray}
|\tilde{\psi}_{k}(p)|^2&=&\frac{1}{\pi L}
[\frac{\sin^2\frac{(p-k)L}{2}}{(p-k)^2}
+\frac{\sin^2\frac{(p+k)L}{2}}{(p+k)^2}
\nonumber\\
&&
+\frac{2\sin(\frac{(k-p)L}{2})
2\sin(\frac{(k-p)L}{2})}
{(p-k)(p+k)}\cos(kL)].
\end{eqnarray}

	On the other hand, if we recall the definition of the delta 
function
\begin{equation}
\delta(x)=\lim_{L\rightarrow \infty}\frac{1}{\pi}\frac{\sin(x\cdot 
L)}{x},
\end{equation}
it is easily verified that, when $L\rightarrow \infty$,
we have
\begin{equation}
P_1(p)=\frac{L}{2\pi}N_p=
\frac{L}{2\pi}\frac{1}{\exp(\frac{E_p-\mu}{T})-Q}.
\label{modifiedbe}\end{equation}

	That is, in the limit of an infinite 1-D box, we obtain a modified 
Bose-Einstein distribution, where the deformation parameter $Q$ 
replaces 
the unity factor, characteristic of BE statistics. In the finite box 
case,
however, the spectrum should change more drastically, due to quantum 
effects, 
which we already showed in Ref.\cite{our1}. Moreover, in the present 
case, 
we have both the finite size and the deformation parameter effects 
combined. To illustrate this, we show in Fig.1, the single spectrum 
distribution for two different box sizes. In that plot, as in all the 
others that follow, we have chosen a null chemical potential, 
i.e., $\mu = 0$, for simplicity. For comparison, the Bose-Einstein 
spectrum distribution, as well as the corresponding modified form given 
by Eq. (\ref{modifiedbe}), are also shown. 
It is interesting to note that for finite systems and decreasing values 
of the $Q$ parameter, the width of single $Q$-boson distribution 
becomes broader, causing the maximum of the distribution to drop and 
the 
tail to rise, due to the conservation of the number of particles. The 
drop of the maximum for the same value of the momentum but for 
a smaller value of Q would correspond to a {\it weaker bosonic} 
behavior of the particles when compared to the $Q \longrightarrow 1$ 
limit, leading to a lower occupancy for small values of the momenta. 
The effect is more pronounced for 
increasing size of the emission region. On the other hand, decreasing 
the values of the deformation parameter $Q$ has a similar effect as to 
decreasing the source emission size (see Ref.\cite{our1}), which is 
consistent with the uncertainty principle since, as the volume of the 
system decreases, the uncertainty in the pion coordinate decreases 
accordingly, causing larger fluctuations in the pion momentum 
distribution, 
which then becomes broader. 

	It is interesting to check how our result would compare with 
the interpretation given in Ref.\cite{gastao}, for which $Q$ could be 
viewed as an effective parameter reflecting the internal degrees of 
freedom of the bosons. In that reference, the deformation parameter $Q$ 
is 
related to the ratio of the bosonic volume ($L^3$) to the system volume 
($V$) by $Q^2 \approx 1 - L^3/V$, where $L$ is the boson's RMS radius. 
The ratio is then correlated to the degree of {\sl bosonic overlap}. 
Although we do not consider here the bosons as extended objects, 
we still could try and see if that picture is compatible with our 
study. 
Let us first consider that the bosons have a fixed size.  
We then compare the above relation 
for two values of the system volume (where $V_2>V_1$), associating a 
value of $Q$ to each case. It is very simple to see that 
$Q_2 = \sqrt{1-\frac{V_1}{V_2}(1-Q_1^2)}$, i.e., an increase in the 
volume 
would result in a smaller deformation parameter, reflecting a smaller 
overlapping of the bosons and their constituents. In other words, for a 
fixed bosonic size and if we enlarge the volume that contains the 
bosons, the 
{\it resolution} decreases, implying that $Q$ increases, i.e., gets 
closer 
to the boson statistics case for which $Q=1$. Let us take another 
approach, 
by considering $Q$ fixed and studying what happens for increasing 
volumes. 
In this case, a system of $Q$-bosons in a volume $V_1$ would be 
associated to a
$L_1^3/V_1$ and another one, in similar conditions but with a 
volume $V_2>V_1$, would have $L_2^3/V_2$. In order to keep 
$Q$ the same, the ratio has to be kept the same, which means that 
$L_2>L_1$. This could be interpreted as if we had  higher resolution of  
the internal degrees of freedom in the second case (i.e., the boson 
with 
bigger $L$ would have the effect of its internal constituents more 
sharply 
probed). Consequently, 
for the same value of $Q$, we would expect that the larger the system 
is, 
more sensitive it would be to fixed value of Q. This is precisely what 
we can see in Fig.1, since the effect is more pronounced for $L=8$ fm 
than it is for $L=4$ fm. 

	We have seen that the deformation parameter has a significant effect 
on the spectrum of the bosons. We discuss next what this implies to the 
interferometry of two-identical $Q$-bosons. 
In Fig. 2, the correlation functions for two values of the 
mean momentum $K$ are shown for different deformed parameter, $Q$,
as a function of the pair relative momentum, $|{\bf q}|$. For $Q = 1$, 
as 
already shown in Ref.\cite{our1}, we see that, as the mean momentum 
increases, 
the source radius increases accordingly, due to the fact that 
contributions 
from small momenta come from smaller quantum states $l$ which, in turn, 
corresponds to larger spread in coordinate space. 
Similar behavior in the radius is seen as $Q$ decreases below unity.

Another interesting point concerns the way 
the parameter $Q$ changes the intercept 
parameter, $\lambda$, 
of the two-$Q$-boson correlation function. 
The effects on $\lambda$ are more pronounced as smaller values 
of $K$ are considered, which is natural, as for large values of the 
average 
momentum, the quantum effects become less relevant. From 
Eq.(\ref{chaot}), 
we can see that, for increasing $K$, the dominant factors  
come from the larger $l$ states which, on the other hand, give smaller 
contribution to the two-$Q$-boson correlation, due to the 
factor $N_\lambda$, which decreases for increasing $K$, as can be seen 
from Eq. (\ref{nlambda}). Consequently, this makes the intercept 
parameter to vary more slowly for increasing values 
of $K$. This is illustrated in Fig. 3, where $\lambda$ is shown as a 
function
of $Q$ for different values of the mean momentum and for different 
source radii. Again, we note that as the source radius becomes
bigger, the $Q$ effects on the intercept parameter become 
less significant, since in this case the quantum effects are smaller.
In the plot, we only shown the variation of $\lambda$ for $Q$ in the 
interval $[0,1]$, corresponding to $\lambda \le 1$ . Of course, if $Q$ 
is larger than one, as one could expected from Eq.(\ref{chaot}), the 
value of $\lambda$ could be bigger than one.
Also if the value of $Q$ is negative, $\lambda$ could be less than 
zero. 
However, we are treating here bosons with a modified commutation 
relation. 
Since such unexpected behavior for the 
intercept parameter was never observed
experimentally, for any type of bosons, we do not consider 
this case here. 
In other words, our analysis refers basically to the 
interval, $0 \le Q \le 1$. Nevertheless, we should keep in mind that 
the intercept parameter of the two identically charged 
$Q$-bosons could be bigger than one or less than zero, for some 
specific values of $Q$.

Although not shown in Fig. 3, the limit $Q \rightarrow 0$ deserves a 
closer 
analysis. As we briefly discussed in Section II, { the naive classical 
limit $C_2({\bf p_1,p_2}) \rightarrow 1$ is recovered only for single modes}. 
The multi-mode case is analyzed in detail in Appendix I 
but we summarize the main results here. 
We start with the limit $Q \rightarrow 0$, for which 
the first of the commutation relations for equal modes in 
Eq.(\ref{commutators}) is reduced 
to $a a^+ = 1$. Nevertheless, we demonstrate in the Appendix I that  
it is not only the commutation relations that matter 
when discussing the behavior of the intercept parameter, $\lambda$. 
The density matrix seems to play an essential role, 
at least for the type of Q-boson we analyze here. 
In this case, with the density as defined in 
Eq.(\ref{HN}), we get $\lambda \rightarrow 0$ for very small system sizes  
(and $Q=0$), recovering what would be naively expected for classical 
particles. In the opposite limit, i.e., for very large systems, 
we get a constant $N_k$, resulting in $\lambda \rightarrow 1/3$. 

The above limits could suggest that we get different results depending 
on the wave function and boundary conditions, reflecting the 
dependence on the dimensions of the system and, consequently, 
on $E_K$, since the particular $\rho_k$ we chose contains 
an explicit dependence on the energy of the state. However, we 
also demonstrate in Appendix I that, if we had chosen a different 
density matrix than the one in Eq.(\ref{HN}), for instance, 
$\rho_k =$ const., the wave function would not play any role, 
since we have $N_k=$const., and there is no energy dependence in this 
case. As a result, we get  $\lambda \rightarrow 1/3$, independently 
on the system size and $E_K$. 
Although we do not show in Fig. 3 the limit $\lambda(K=0)$, we can 
still check the consistency of the results plotted there with the 
analysis 
we have just made. For that, we should look into the smallest 
value of momentum shown in that plot, i.e., $K=0.3$ GeV/c. 
Then, according to our analysis 
in the Appendix I, the value of the intercept parameter for small 
systems would tend to approach the limit $\lambda \rightarrow 0$, 
whereas for large ones, it should approach $\lambda \rightarrow 1/3$. 
Adapted to our plot, this would mean that 
$\lambda(L=4 fm) < \lambda(L=8 fm)$  for $Q=0$ and $K=0.3$ GeV/c, 
which is precisely what is seen in Fig. 3. We can also verify that 
this result is more general, since the same feature is again reproduced 
in Fig. 6, as we shall see in the next subsection. 

Furthermore, if we return to the expression of the generalized Wigner 
function, Eq. (\ref{wigner}), we see that the usual Wigner function 
is recovered (last term vanishes) for $Q=0$ but, again, for single 
modes 
only. For multi-modes, however, the Wigner function is modified, 
even for $Q=0$. This result seems to indicate the existence of some 
kind of {\sl residual correlations}  among the particles in the system 
i.e., 
an extra communication among the particles of different states, beside 
the commutation relation defined among particles in the same state. 

Along the lines discussed above, only recently we became aware of a 
Monte Carlo event generator by Wilk et al., which makes an attempt to 
improve Bose-Einstein correlations in numerical modeling. It would be 
very interesting to run their simulation and check for consistency 
with our numerical calculation of the correlation function, 
$C_2(p_1,p_2)$ and of the intercept parameter, $\lambda(K)$, 
mainly regarding the multi-particle effects.

\subsection{$Q$-bosons are confined inside a sphere }

In this section, we consider the that the pions produced in high 
energy heavy-ion collisions, treated here as the hypothetical 
$Q$-bosons, 
could be bounded in a sphere up to the time immediately preceding the 
freeze-out of the system. This is conceived in such a way that their 
distribution functions are 
essentially the ones they had while confined. Analogously to the 
procedure 
developed in \cite{our1}, the pion wave function in this case should be 
determined by the solution of the Klein-Gordon equation, i.e., 
\begin{equation}
\left [ \Delta+k^2\right ]
\psi({\bf r})=0
\; \;, \label{kg}\end{equation}
where $k^2=E^2-m^2$ is the momentum of the pion. On writing the above
equation, we have assumed confinement, i.e., the potential felt by the 
pion inside the sphere is zero, while outside it is infinite. The
boundary condition to be respected by the solution is
\begin{equation}
\psi({\bf r})|_{r=R}=0
\; \; ,\label{genbc}\end{equation}
where $R$ is the radius of the sphere at freeze-out  time.

The normalized wave function corresponding to the solution of the
above equation can easily be written as

\begin{eqnarray}
\psi_{klm}({\bf r})&=&
 \frac{1}{RJ_{l+\frac{3}{2}}(kR)} \sqrt{\frac{2}{r}} 
Y_{lm}(\theta,\phi)
 J_{l+\frac{1}{2}}(kr)~~~~ (r < R),
\nonumber\\
&=& 0 ~~~~~~~~~~~~~~~~~~~~~~~~~~~~~~~~~~~~~~~~~~~~~~~ (r \geq R).
\nonumber\\
\label{psiklm1}\end{eqnarray}

The momentum of the bounded particle, $k$, can be determined as the
solution of the equation 
\begin{equation}
J_{l+\frac{1}{2}}(kR)=0
\; \;. \label{jlbc}\end{equation}

Inserting Eq. (\ref{psiklm1}) into Eq. (\ref{Fourier}),
we can determine the Fourier transform of
the confined solution of a pion inside the sphere, as a function of the
momentum {\bf p}, as \cite{our1} 

\begin{equation}
\tilde{\psi}_{klm}({\bf p})=\sqrt{\frac{2}{p}} i^l Y_{lm}(\hat{p})
\left[ \frac{-k}{p^2-k^2} \right] J_{l+\frac{1}{2}}(pR)
\; \;. \label{psiklm3}\end{equation}

In terms of Eq. (\ref{Nlamb}) and (\ref{p1g}), 
the single-inclusive distribution function can be written as 

\begin{eqnarray}
P_1({\bf p})&=&\sum_{klm}N_{klm}
\tilde{\psi}_{klm}^*({\bf p})\tilde{\psi}_{klm}({\bf p})
\nonumber\\
&=&
\sum_{k,l}\frac{1}{\exp{(\frac{E_{kl}-\mu}{T})}-Q}
\left(\frac{2l+1}{2\pi p}\right)
\left( \frac{kJ_{l+\frac{1}{2}}(pR)}{p^2-k^2} \right)^2,
\nonumber\\
\label{P1sph1}\end{eqnarray}

In the limit $R \rightarrow \infty$, the single particle spectrum,
can then written as
\begin{equation}
P_1({\bf 
p})=\frac{1}{\exp{(\frac{E_p-\mu}{T})}-Q}\left[\frac{V}{(2\pi)^3}
\right] \; \;, \label{P1Rinf1}\end{equation}
where $V=\frac{4\pi}{3}R^3$ is the volume of the sphere. We see
from    Eq. (\ref{P1Rinf1}) that
the modified Bose-Einstein distribution is recovered in the limit of 
very large volumes.

In Fig. 4, the normalized single-particle distribution is plotted
as a function of $|{\bf p}|$.
We clearly see that, due to the boundary effects, the
maximum value of $|{\bf p}|$ in the spectrum decreases for decreasing
volumes, being
always smaller than the case corresponding to the $R \rightarrow 
\infty$ limit.
On the other hand, this is similar to the result obtained with the 
previous 
example of the 1-D box, and 
the confinement does not seem to cause a significant effect on the 
spectrum.  

We can write the expectation value of the product of
two $Q$-boson creation operators in momentum space as before, resulting 
in 

\begin{eqnarray}
&&\langle \hat{\psi}^{\dagger}({\bf p_1})\hat{\psi}({\bf p_2})\rangle
=\sum_{klm}\frac{\tilde{\psi}^*_{klm}({\bf p_1})
\tilde{\psi}_{klm}({\bf p_2})}
{\exp{\left( \frac{E_{kl}-\mu}{T} \right)}-Q}
\nonumber\\
&=&
\sum_{klm}\frac{1}{\exp{\left( \frac{E_{kl}-\mu}{T} \right)}-Q}  \times
\nonumber\\
&&\sqrt{\frac{2}{p_1}}
(-i)^l Y_{lm}^*(\hat{p}_1) \left[\frac{-k}{p_1^2-k^2}\right]
J_{l+\frac{1}{2}}(p_1R)
\nonumber\\
&&\sqrt{\frac{2}{p_2}}
(i)^l Y_{lm}^*(\hat{p}_2)\left[\frac{-k}{p_2^2-k^2}\right]
J_{l+\frac{1}{2}}(p_2R)
\nonumber\\
&=&  \sum_{kl}\frac{1}{\exp{\left( \frac{E_{kl}-\mu}{T} \right)}-Q}
\sqrt{\frac{4}{p_1p_2}}
\nonumber\\
&&
\frac{k^2}{(p_1^2-k^2)(p_2^2-k^2)}J_{l+\frac{1}{2}}(p_1R)
\nonumber\\
&&J_{l+\frac{1}{2}}(p_2R)
\left(\frac{2l+1}{4\pi}\right)P_l(\hat{p}_1\cdot\hat{p}_2)
\; \;. \label{psi*psi}\end{eqnarray}

The two-pion interferometry correlation function can then be estimated
by inserting the expressions (\ref{P1sph1}) and (\ref{psi*psi}), into 
Eq. (\ref{c21st}). We see from the above results that, in general, 
this function depends on
the angle between ${\bf p_1}$ and ${\bf p_2}$, similarly to what 
was discussed in \cite{our1}. For the sake of simplicity, however, 
we will also consider here ${\bf p_1}$ parallel to ${\bf p_2}$, 
implying that
$P_l(\hat{{\bf p}}_1\cdot\hat{{\bf p}}_2=\pm 1)=(\pm 1)^{l}$. 
The results for two-pion interferometry corresponding to two 
different values of the pair average momentum ${\bf K=(p_1+p_2)}/2$, 
but fixed temperature, are shown in Fig. 5. For $Q=1$ 
we can see that, as the pair average momentum, $K$,  increases, 
the apparent source radius becomes bigger, which reproduces 
the result obtained in Ref. \cite{our1}. However, considering $K=0.3$ 
GeV/c, 
if we compare the cases corresponding to $Q=1$ and $Q=0.5$, 
respectively, 
we see that the resulting correlation 
function becomes narrower and the intercept parameter, $\lambda$, 
drops below its previous unit value. On the other hand, 
if we now keep this value of $Q=0.5$ but consider 
$K=0.5$ GeV/c, the width is even narrower, but the 
intercept is higher than that corresponding to $K=0.3$ GeV/c, 
although it still is below one. 
This due to the fact that for smaller momentum pairs, the
quantum effects are stronger, as in the previous toy model 
studied in Fig. 2. The effect comes from the contribution of the third term 
in 
two-$Q$-boson interferometry formula, in Eq. (\ref{c21st}). 
%(\ref{c23rd}). 
We could understand these results by noting that
pions with larger momentum come
from larger quantum $l$ states which, in turn, correspond to a
smaller spread in coordinate space. Due to the weight factor in
Eq. (\ref{c21st}), of modified Bose-Einstein form,
larger quantum states will give a smaller contribution
to the source distribution, causing the effective
source radius to appear larger. On the other hand, this 
behavior is interesting if we compare to results corresponding to 
expanding systems. In this last case, the probed part of the system 
decreases with increasing average momentum\cite{UH,Csorgo}.
Naturally, our present approach does not
consider the effects of expansion and the enlargement of the system's
apparent dimensions with increasing K, seen in Figure 5, has its origin
in the strong sensitivity to the dynamical matrix.
In Ref.\cite{AS01}, the combined effects of a finite boundary 
and an expanding system were considered together. What they 
observed was an opposite effect, i.e., the effective source radius 
extracted from two-pion interferometry would decrease 
as $K$ increases. However, for small values of the momentum $K$, it 
seems
that the boundary effects are dominant over the expansion effects. 
 
In Fig. 6, we plot the intercept parameter 
$\lambda$ vs. $Q$ for different values of momenta and source
radii. We see that the similarity to the results in Fig. 3 is evident: 
$\lambda$ becomes bigger as either the total
momentum of the pair, $K$, or the source radii, $R$, increases, 
reaching values significantly below one for sufficiently small values 
of either one of those variables. 
This is expected, since the quantum effects are more prominent for 
smaller momenta. Again, the comments relating these results to 
the speculation in Ref.\cite{gastao} apply here, as in the 
previous example. 

   In summary, by comparing the interferometric results  
corresponding to the two toy models, we see that a 
general behavior is roughly reproduced in both cases. 
First, Figures 1 and 4 show that the results for the normalized 
spectra are very similar, the maximum of the curves dropping with 
decreasing values of $Q$, followed by the rise of the respective tails, 
this being related to the conservation of the number of particles. 
Second, the width of the correlation functions, as seen in Figures 2 and 5, 
seems to decrease for increasing values of $Q$. 
Besides, we see from the Figures 3 and 6 that the $Q$-boson effects 
cause the intercept parameter, $\lambda$, 
to drop below unity. This clearly demonstrates that 
treating pions as $Q$-boson indeed alter the two-pion interferometry. 
Besides, we saw in our examples that it also modifies and 
generalizes the form of two particles Wigner function. 
Nevertheless, we also saw that the  
effects for the bounded case are less pronounced as compared to 
the unbounded one, at least for the set of parameters adopted in 
the calculation.

\section{Conclusions}

In this paper, we derive spectrum and correlation function relations 
by adopting the density matrix given in Eq.(\ref{rho}), suitable 
for describing charged $Q$-bosons. 
The finite volume effects on the $Q$-boson spectrum were then
studied in Figures 1 and 4, for two specific examples, leading to 
similar 
results as in Ref.\cite{our1,MW93,MW95,AS97}, for $Q = 1$. 
We find that the small momentum
region is depleted for the modified Bose-Einstein distribution 
with respect to case when $Q \rightarrow 1$.
The effects on two equally charged $Q$-boson correlation function were 
also
analyzed here. The results in Figures 2 and 5 show that the correlation
function shrinks for increasing average pair momentum, corresponding to
an increase of its inverse width\cite{our1}. We also observe that 
its intercept drops for decreasing $Q$. In other words, it was shown 
in Figures 3 and 6 that the intercept parameter, $\lambda$, decreases 
for increasing $Q$, for the same values of the average transverse 
momentum and source radius. On the other hand, $\lambda$
becomes larger when either the 
mean momentum of two-$Q$-boson or the source radius increases. 
This  result reflects a strong sensitivity to the dynamical matrix,
through the modified Bose-Einstein weight factor.
For $Q=1$, previous pion interferometry results are regained, 
as well as the Boltzmann-type distribution for $Q=0$, as can be 
verified in Ref. \cite{our1}.

	We have also derived a generalized version of the two-boson 
Wigner function, allowing to treat the case of two $Q$-bosons. 
This result is particularly important because it shows that, 
in more general situations, the Wigner function is distorted with 
respect to its decoupled form, i.e., $g ({\bf x},{\bf K} ; 
{\bf y},{\bf K}) = g({\bf x},{\bf K}) g({\bf y},{\bf K})$. 
The extra terms it acquires, as shown in Eq. (\ref{wigner}) (see also 
Eq. (\ref{wignerB}) in the Appendix II), 
reflect non-trivial interactions among the $Q$-bosons. This 
generalization, however, is reduced to its well-known form in the 
limit $Q \rightarrow 1$, as it should. 

	We also analyzed in the current work how our results compare with 
the interpretation given in Ref.\cite{gastao}, for which $Q$ could be 
viewed as an effective parameter reflecting the internal degrees of 
freedom of the bosons. If we consider a fixed value for 
the deformation parameter, $Q$, and compare results for increasing 
volumes, our result would be compatible with that interpretation. 
To summarize the comparison we could say that, for increasing volumes 
and keeping the same value 
of $Q$, we would have to increase the {\it resolution}, i.e., the 
sensitivity 
of the probe to the internal degrees of freedom of the boson. 
Consequently, 
for the same value of $Q$, we would expect that the larger the system 
is, 
more sensitive it would be to this parameter. This is precisely what 
we can see in Fig.1, since the effect is more pronounced for $L=8$fm 
than it is for $L=4$fm. 

	The derivation analyzed here has several common features with 
the one in Ref.\cite{agp,AGI} but here we adopt a entirely different 
approach, focusing in what seems to us the most important part of an 
interferometric analysis, i.e., the correlation functions themselves. 
We also analyze the effects on the spectra and on the intercept  
parameter, $\lambda$, but, 
again, our result is more general than in that reference, since it is 
there restricted to single modes. 

Another remark concerns the relation of the parameter 
$\lambda$ and its possible interpretation as a partial 
coherence of the emitting source for values below unity, as commonly 
found in the literature on boson interferometry. 
It is well-known that many effects, such as resonances, 
dynamical and multi-particle effects, as well as kinematical cuts, 
could also contribute to yield values of so-called chaoticity or 
coherence parameter below unity, although those effects have no 
relation to partial coherence of the source. Similarly, the behavior we 
discussed of the intercept parameter $\lambda$, when considering 
decreasing values of the deformation $Q$, are not related to the way 
the source emit those bosons. The effects discussed here are meant to 
show that they also could cause a deviation from the idealized picture. 
Moreover, for keeping the analysis simple, we are not taking all the above 
effects into account in our study. For this reason, we preferred to 
not compare our results with experimental data, and thus did not 
introduce any quantitative analysis of this picture.

Also, for completeness, we derive 
in the Appendix II, the equations  
associated to the so-called {\it type-B $Q$-boson 
interferometry}\cite{agp,AGI}, corresponding to slightly different 
commutation relations. The derived relations for the single-  
and two-particle distributions are in 
Eq.(\ref{P1B}), and Eq. (\ref{P2B}), respectively. 
Also for this case, we propose a 
generalized form for the Wigner function, as can be seen in 
Eq.(\ref{wignerB}).

\acknowledgments

S.S.P. is deeply grateful to Keith Ellis and 
the Theoretical Physics Department 
at Fermilab, as well as to Larry McLerran and 
the Nuclear Theory Group at BNL, for their kind hospitality when 
finalizing 
this work. 
This research was partially supported by CNPq (Proc. N. 200410/82-2). 
This manuscript has been authored under Contracts No. DE-AC02-98CH10886 
and No. DE-AC02-76CH0300 with the U.S. Department of Energy. 
This work was also partially supported 
by Funda\c{c}\~ao de Amparo \`a Pesquisa do Estado de S\~ao Paulo 
(FAPESP, Proc. N. 1998/05340-2 and 1998/2249-4), Brazil, and 
NSERC of Canada. The authors would like to 
thank Charles Gale for interesting discussions and 
Gast\~ao Krein for his attentive reading 
and comments about the manuscript. We would also like to thank 
G. Wilk for asking us interesting questions, whose answers 
lead us to include an extra appendix to the manuscript.

%%%%%%%%%%%%%%%%%%%%%%%%%%%%%%  FIGURES  %%%%%%%%%%%%%%%%%%%%%%%%%%%%%%

\vskip 1.8cm
\begin{figure}[h]\epsfxsize=8cm
\centerline{\epsfbox{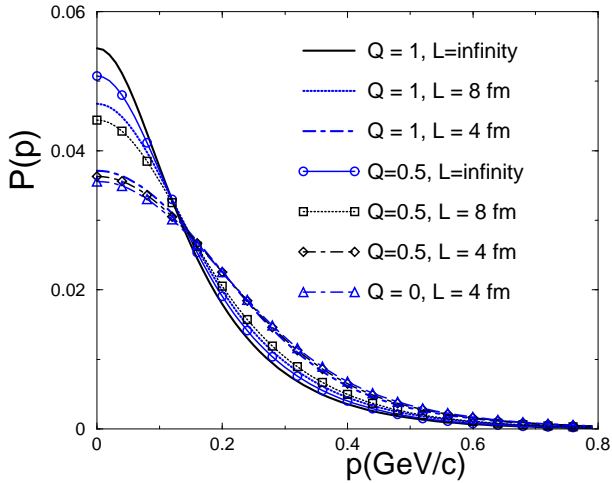}}
\caption{(Color online) The normalized spectrum (in arbitrary units)
vs. momentum $|${\bf p}$|$ (in GeV/c) is shown.
The input temperature is $T=0.14$ GeV and the chemical potential is
$\mu=0$. The solid line
corresponds to the modified Bose-Einstein distribution, i.e., to the 
limit
$R \rightarrow \infty$. The dotted and dashed lines correspond, 
respectively,
to $L=8$ fm and $L=4$ fm. The thicker lines refer 
to $Q=1.0$ and the thinner ones to $Q=0.5$.}
\end{figure}

\begin{figure}[h]\epsfxsize=8cm
\centerline{\epsfbox{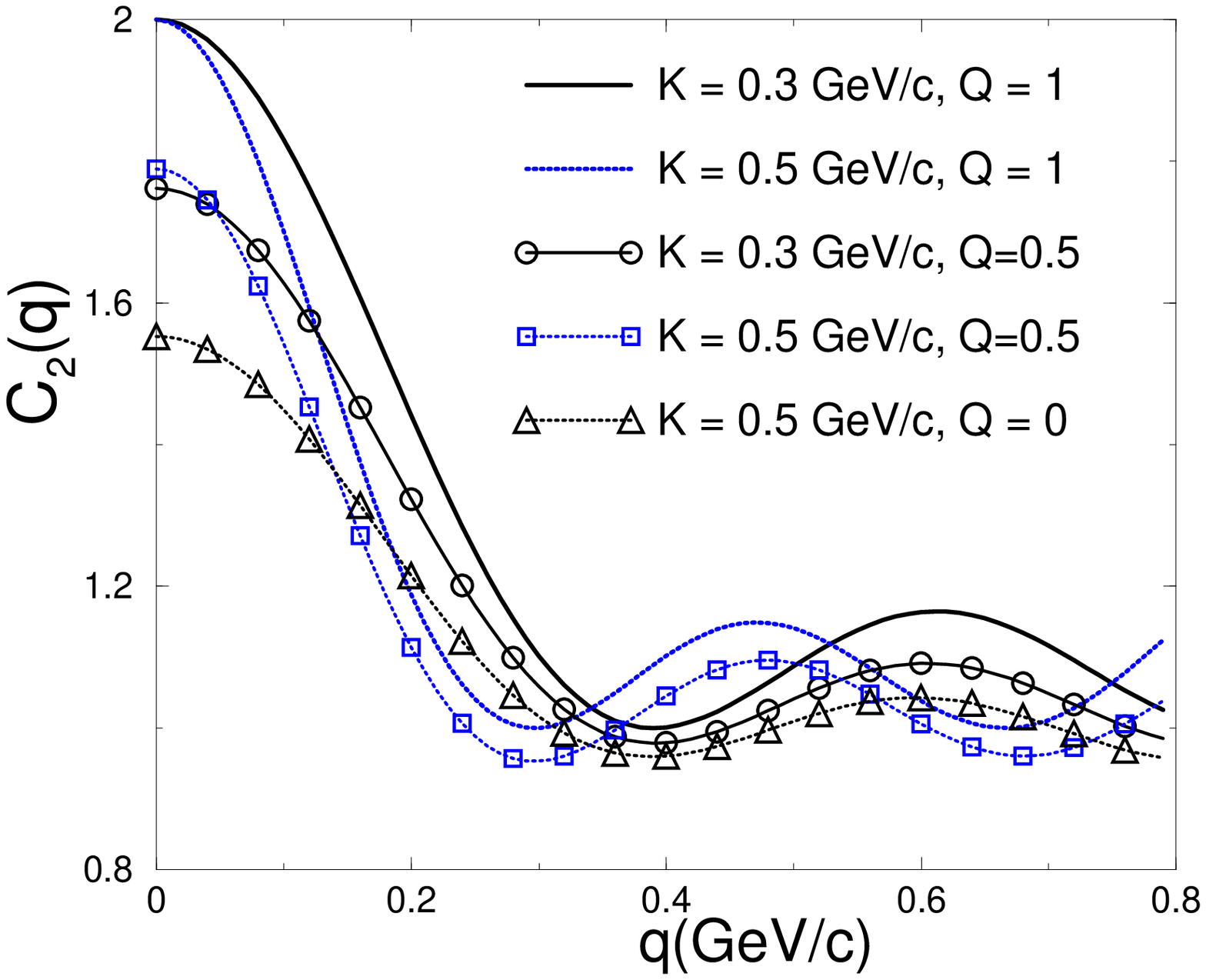}}
\caption{(Color online) Two-pion correlation
vs. momentum difference $|${\bf q}$|$ (in GeV/c).
The input temperature is $T=0.14$ GeV and the chemical potential is
$\mu=0$. The solid line corresponds to 
mean momentum $K=0.3$ GeV/c and dashed one to $K=0.5$ GeV/c.
The thicker lines refer  
to the case of $Q=1.0$ and the thinner one to
$Q=0.5$. The box size is $L= 4$ fm.}
\end{figure}

\begin{figure}[h]\epsfxsize=8cm
\centerline{\epsfbox{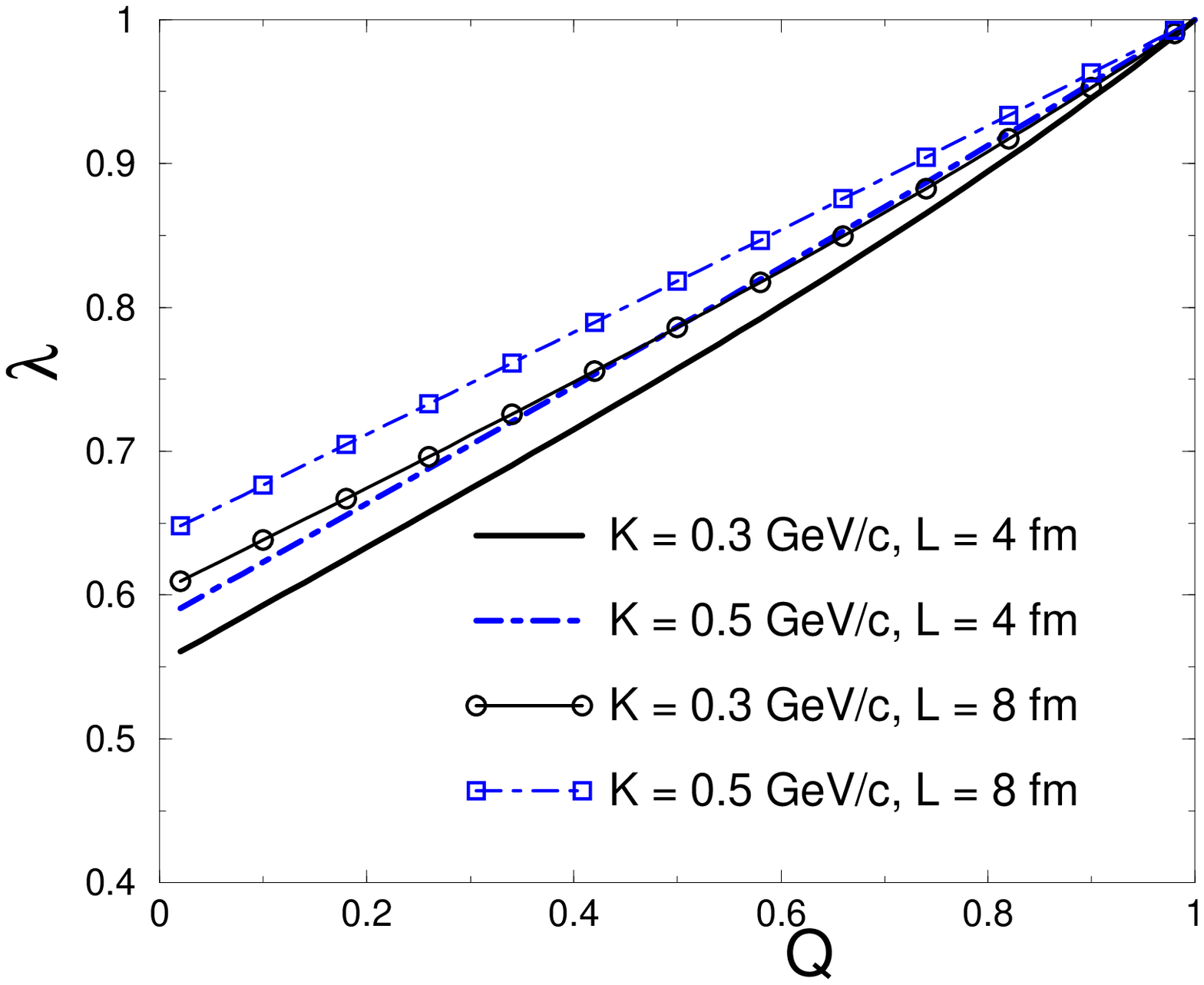}}
\caption{(Color online) The intercept parameter, $\lambda$, is shown 
vs. $Q$, the deformation parameter.
The input temperature is $T=0.14$ GeV and the chemical potential is
$\mu=0$. The solid line 
corresponds to mean momentum $K=0.3$ GeV/c and dashed one 
to the case $K=0.5$ GeV/c.
The thicker lines refer 
to the case $L=4$ fm and the thinner ones to $L =8$ fm.}
\end{figure}

\begin{figure}[h]\epsfxsize=8cm
\centerline{\epsfbox{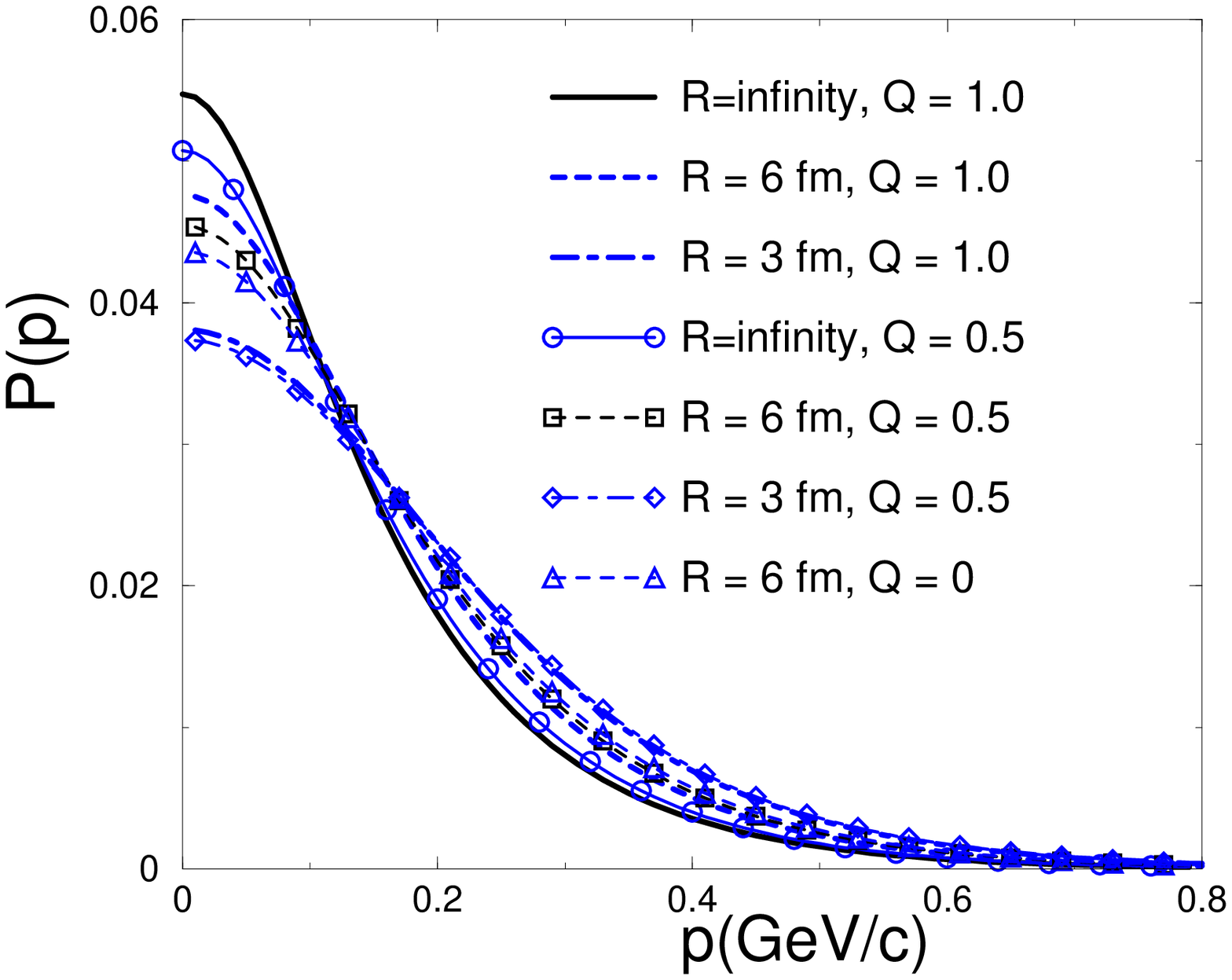}}
\caption{(Color online) The normalized spectrum (in arbitrary units)
vs. momentum $|${\bf p}$|$ (in GeV/c) is shown.
The input temperature is $T=0.12$ GeV and the chemical potential is
$\mu=0$. The solid line corresponds to the case $R=3$ fm, the  
dotted one to the case $R= 6$ fm, and 
the dashed line corresponds to the case, $R=\infty$.
The thicker lines refer to $Q=1.0$ and the thinner ones to$Q =0.5$.}

\begin{figure}[h]\epsfxsize=8cm
\centerline{\epsfbox{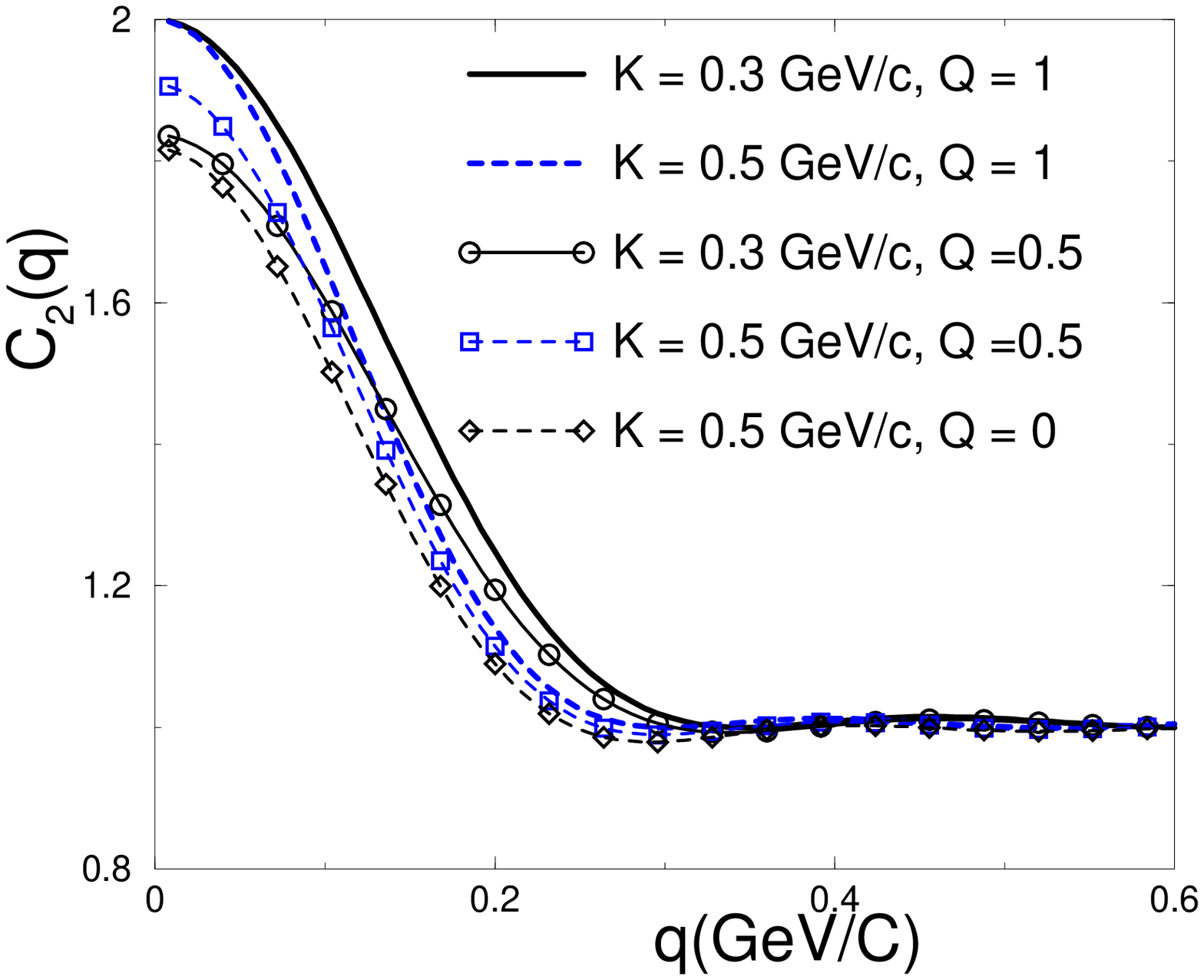}}
\caption{(Color online) Two-pion correlation
vs. momentum difference $|{\bf q}|$ (in GeV/c) is shown.
The input temperature is $T=0.12$ GeV and the chemical potential is
$\mu=0$. The solid line
corresponds to mean momentum $K=0.3$ GeV/c and the dashed one, 
to the case $K=0.5$ GeV/c.
The thicker lines refer 
to the case of $Q=1.0$ and the thinner ones to
$Q=0.5$. The sphere size is $R= 3$ fm.}
\end{figure}

\begin{figure}[h]\epsfxsize=8cm
\centerline{\epsfbox{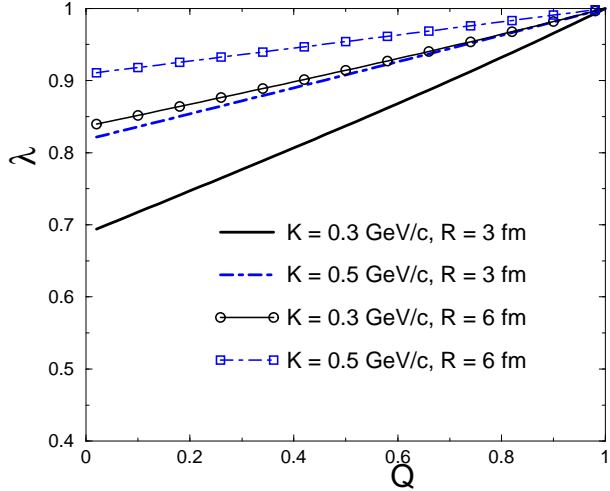}}
\caption{(Color online) The intercept parameter, $\lambda$, is shown 
vs. the deformation $Q$.
The input temperature is $T=0.12$ GeV and the chemical potential is
$\mu=0$. The solid line
corresponds to mean momentum $K=0.3$ GeV/c and dashed one 
to the case $K=0.5$ GeV/c.
The thicker lines refer 
to the case $R=3$ fm and the thinner ones to
$R =6$ fm.}

\end{figure}
\end{figure}

\vfill
\newpage

\vskip2cm
\begin{center}
{\bf APPENDIX I}
\end{center}
\vskip0.5cm

	We here analyze the $Q=0$ limit in more detail. We can see its 
implications on the expectation values, directly from 
Eq.(\ref{a+a})-(\ref{a+a+aa}), on the correlation function, from 
Eq.(\ref{c21st}), and on the intercept parameter, $\lambda$, from 
Eq.(\ref{chaot}), by simply imposing the limit $Q=0$. 
Alternatively, we can start with our definitions in 
Eq.(\ref{commutators}), 
and thus simultaneously test our results.  
Lets also go back to the definition in 
Eq.(\ref{nstate}). For Q=0 we see that 
\begin{equation}
[n]=  \; \; \left\{ \begin{array}{l}
            \mbox{ 1 $\rightarrow $  $ n \ne 0$} \; \\
	    \mbox{ 0 $\rightarrow $  $ n = 0$} 
             \end{array}
             \right\}\; . 
\label{n1}\end{equation}
Thus, $[n]! = 1$ in both the above cases. Consequently, instead of 
Eq.(\ref{nstate}), we have 
\begin{equation}
| n \rangle_{_l} = (a_{\l}^{\dagger})^n |0 \rangle \; . 
\label{nstate-Q0}\end{equation}

If we apply to the above equation the annihilation and creation 
operators, we have
\begin{equation}
a_{\l} | n \rangle_{_l} = a_{\l}(a_{\l}^{\dagger})^n |0 \rangle = |n-1 
\rangle 
\; ; \; a_{\l}^{\dagger} | n \rangle_{_l} = |n+1 \rangle \; ,  
\label{a-to-n}\end{equation}
which follow from the first commutation relations in 
Eq. (\ref{commutators}) and from the condition $ a_{\l} |0 \rangle = 
0$. 
The other commutation relations remain unchanged in this limit.

Let us now estimate the trace of the density matrix, $tr(\rho_l)$. 
From Eq.(\ref{rho}) and (\ref{HN}), we have 
\begin{eqnarray}
tr(\rho_{l})&=& \sum_{n} {_{_l}\!\langle} n| \rho_l | n \rangle_{_l} = 
\sum_{n=0}^\infty \sum_{m=0}^\infty [-\frac{1}{T}(E_l-\mu)n]^m 
\frac{1}{m!} = 
\nonumber \\
&=&
\sum_{n=0}^\infty e^{-\frac{1}{T}(E_l-\mu)n} = 
\frac{1}{1-\exp[-\frac{1}{T}(E_l-\mu)]}
\; \;.\label{trace2}\end{eqnarray}

For estimating the limits of 
Eq.(\ref{a+a}) and (\ref{a+a+aa}) 
for $Q\rightarrow 0$ along the lines above, we have to estimate the 
traces 
$tr(a^{\dagger}_{l}a_{l}\rho_{l})$ and 
$tr(a^{\dagger}_{l}a^{\dagger}_{l}a_{l}a_{l}\rho_{l})$, 
as follows 

\begin{eqnarray}
tr(a^{\dagger}_{l}a_{l}\rho_{l}) &=&
\sum_{n} {_{_l}\!\langle} n| a^{\dagger}_{l}a_{l}\rho_l | n 
\rangle_{_l}= 
\nonumber\\
&=& \sum_{n=0}^\infty \sum_{m=0}^\infty {_{_l}\!\langle} n| 
a^{\dagger}_{l}a_{l} 
[-\frac{(E_l-\mu)}{T}]^m \frac{N_l}{m!} | n \rangle_{_l} =
\nonumber\\
&=& \sum_{n=1} \sum_{m} [-\frac{1}{T}(E_l-\mu)n]^m \frac{1}{m!} 
{_l}\!\langle n-1 | n-1 \rangle_{_l} 
\nonumber\\
&=& \sum_{n=0} e^{-\frac{1}{T}(E_l-\mu)n}-1=
\nonumber\\
&=& \frac{e^{-\frac{1}{T}(E_l-\mu)}}
{1-\exp[-\frac{1}{T}(E_l-\mu)]} 
\; \;.\label{a+arho}\end{eqnarray}

\vfill
\newpage

\begin{eqnarray}
&& tr(a^{\dagger}_{l}a^{\dagger}_{l}a_{l}a_{l}\rho_{l})=  
\sum_{n} {_{_l}\!\langle} n| 
a^{\dagger}_{l}a^{\dagger}_{l}a_{l}a_{l}\rho_l 
| n \rangle_{_l} = 
\nonumber\\
&=&
\sum_{n=0}^\infty \sum_{m=0}^\infty {_{_l}\!\langle} n| 
a^{\dagger}_{l}a^{\dagger}_{l}a_{l} a_{l} 
[-\frac{1}{T}(E_l-\mu)]^m \frac{N_l}{m!} | n \rangle_{_l} =
\nonumber\\
&=&
\sum_{n=1}^\infty \sum_{m=0}^\infty 
[-\frac{1}{T}(E_l-\mu)n]^m \frac{1}{m!} {_{_l}\!\langle} n-1| 
a^{\dagger}_{l} a_{l} | n-1 \rangle_{_l} = 
\nonumber\\
&=&
\sum_{n=2}^\infty \sum_{m=0}^\infty [-\frac{1}{T}(E_l-\mu) n]^m 
\frac{1}{m!} 
{_{_l}\!\langle} n-2 | n-2 \rangle_{_l} =
\nonumber\\
&=& 
\sum_{n=0}^\infty 
e^{-\frac{1}{T}(E_l-\mu)n} - e^{-\frac{1}{T}(E_l-\mu)} - 1  = 
\frac{e^{-\frac{2}{T}(E_l-\mu)]}}
{1-e^{-\frac{1}{T}(E_l-\mu)}}
.\nonumber\\
\label{a+a+aarho}\end{eqnarray}

From the above results and our definition in 
Eq.(\ref{A}), it follows immediately that 
\begin{equation}
\langle a_{l}^{\dagger}a_{l}\rangle = e^{-\frac{1}{T}(E_{l}-\mu)} = N_l 
\; ,\label{a+a2}\end{equation}
and 
\begin{equation}
\langle a_{l}^{\dagger}
a_{l}^{\dagger}
a_{l}
a_{l}\rangle
=
e^{-\frac{2}{T}(E_{l}-\mu)}
\;.\label{a+a+aa2}\end{equation}

It is then straightforward to conclude that both results coincide with 
the ones in Eq.(\ref{a+a}) and 
(\ref{a+a+aa}), reinforcing the correctness of our results. 

The purpose of the above derivation goes beyond what we have just 
discussed, it is also helpful for discussing more deeply the peculiar 
limit 
of $Q \rightarrow 0$. As we have already pointed out in the body of the 
manuscript, we recover the expected classical result in that limit for 
single modes only. In other words, only for {\sl single modes} we have, 
for $q=p_1-p_2=0$ (and, consequently, $K=p_1=p_2$)
$$
\lim_{Q \rightarrow 0} C_2(K,K) = 2 \; \; \; ; \; \; \; \lim_{Q 
\rightarrow 0} 
\lambda(K) = 0. 
$$
In this case, we also recover the well-known form of the Wigner 
function, i.e. 
$$
\lim_{Q \rightarrow 0} g_{l} ({\bf x},{\bf K}) = g({\bf x},{\bf K}) 
g({\bf y},{\bf K}).  
$$

Nevertheless, when multi-modes are taken into account, the behavior of 
the above quantities change considerably. For better illustrating this 
fact, 
it is convenient to adopt a specific example. Let us choose our toy 
model in section III.A, for simplicity, restricting ourselves to 
the $Q=0$ case. In general, 
$\lambda = \lambda(q=p_1-p_2=0) = \lambda (K=p_1=p_2)$ but, for 
simplifying our analytical study, 
we restrict our analysis to $\lambda (K=0)$. 
In this case, we see that the square modulus 
of the wave-function in momentum space can be written as 

\begin{equation}
|\tilde{\psi}_{k}(0)|^2=\frac{2\sin^2(kL/2)}{\pi L k^2} 
[1 - \cos(kL)] ; kL=n\pi (n=1,2...),
\label{psi1}\end{equation}
from which we see that only odd values of $n$, i.e., $n=2m-1$ can 
contribute, resulting in 

\begin{eqnarray}
|\tilde{\psi}_{k}(0)|^2  =   
\left\{ \begin{array}{l}
            \mbox{   $ \frac{4 L}{(2m-1)^2 \pi^3}$ (m=1,2,...)} \; \\
	    \mbox{   } \; \\
	    \mbox{   ~ 0 ~(n even)} 
             \end{array}
\right .
%             \right\}   
\end{eqnarray}

Then, for writing the chaoticity parameter in the limit of $Q=0$ for 
the 
toy model, we rewrite 
Eq.(\ref{chaot}) for this particular case, as 

\begin{equation}
\lambda({\bf K}) = 1 -
\frac{ \sum_{l} \; N_{l}^2 \;
|\tilde{\psi}_{l}({\bf K})|^4}
%%%%\exp(-\frac{2E_{l}}{T})}
{(\sum_{l}N_{l}|\tilde{\psi}_{l}({\bf K})|^2)
(\sum_{l}N_{l}|\tilde{\psi}_{l}({\bf K})|^2)}
\; . \label{chaotoy} 
\end{equation}

We have then to estimate the sums separately, as follows

\begin{equation}
\sum_{l} N_{l} |\tilde{\psi}_{l}({\bf K})|^2 = 
\sum_{l} e^{-\frac{E_{l}}{T}} |\tilde{\psi}_{l}({\bf K})|^2
\; . \label{sum2} 
\end{equation}

\begin{equation}
\sum_{l} \; N_{l}^2 \;
|\tilde{\psi}_{l}({\bf K})|^4 = 
\sum_{l} \; e^{-\frac{2E_{l}}{T}} \;
|\tilde{\psi}_{l}({\bf K})|^4 
\; . \label{sum4} 
\end{equation}

For estimating the above sums, it is more convenient to consider 
appropriate limits for the size of the 1-dimensional box. 
Let us first assume the limit of very small sizes, i.e., 
$L \rightarrow 0$. In this case, $ E_l \approx  n \pi/L $, 
we denote as 

\begin{eqnarray}   
\left\{ \begin{array}{l}
            \mbox{$x = e^{-\frac{E_1}{T}} \ll 1$ ($n=1$)} \; \\
	    \mbox{   } \; \\
	    \mbox{$E_{\tiny 2m-1}=(2m-1)\pi/L = (2m-1)E_1$}
             \end{array} \nonumber
             \right\} \Rightarrow \\
             N_{2m-1} = x^{2m-1} 
%%%%%             \Rightarrow N_{2m-1} = x^{2m-1} 
\end{eqnarray}

Consequently, the sums for ${\bf K} = 0 $ can be written as 

\begin{eqnarray}
\sum_{m} N_{m} |\tilde{\psi}_{m}({\bf 0})|^2 &=& 
\sum_{m} x^{2m-1} \frac{4L}{\pi^3} \frac{1}{(2m-1)^2}=
\nonumber \\ 
&=& \frac{4L}{\pi^3} \sum_{m} \frac{x^{2m-1}}{(2m-1)^2}
\; . \label{sum2a} 
\end{eqnarray}

\begin{eqnarray}
\sum_{m} \; N_{2m-1}^2 \;
|\tilde{\psi}_{2m-1}(0)|^4 &=& 
\sum_{m} \; x^{2(2m-1)} \; \frac{16 L^2}{(2m-1)^4 \pi^6} = 
\nonumber \\ 
&=&  \frac{16 L^2}{\pi^6} \sum_{m} \frac{x^{2(2m-1)}}{(2m-1)^4}
\; . \label{sum4a} 
\end{eqnarray}

If we then bring Eq.(\ref{sum2a}) and Eq. (\ref{sum4a}) into 
Eq. (\ref{chaotoy}), we 
get 
\begin{eqnarray}
&& \lim_{L\rightarrow0 (x\rightarrow 0)} \lambda({\bf K=0}) = 
1 - \frac{\sum_{m} \frac{x^{2(2m-1)}}{(2m-1)^4}}
{\left\{\sum_{m} \frac{x^{2m-1}}{(2m-1)^2}\right\}^2} = 
\nonumber \\
&=& 1 - \lim_{L\rightarrow0 (x\rightarrow 0)} 
\frac{x^2(1+\frac{x^4}{3^4}+\frac{x^8}{5^4}+\frac{x^{12}}{7^4}+ ...)}
{x^2(1+2\frac{x^2}{3^2}+\frac{x^4}{3^4}+2\frac{x^4}{5^2}+ ...)}=1-1=0 
,\nonumber \\
 \label{chaotoy2} 
\end{eqnarray}
and thus, the expected limit for {\sl classical} particles is recovered 
for $Q=0$, when $L \rightarrow 0$. 

In the opposite limit of $L \rightarrow \infty$, however, we can see 
that 
$k=(2m-1)\pi/L \rightarrow 0$, which implies that $E_k \rightarrow 
m_\pi$. 
Consequently, $N_k \rightarrow N_0$ = const. 
In this case
\begin{eqnarray}
\lim_{L \rightarrow 0 (x \rightarrow 0)} \lambda({\bf K=0}) &=& 
1 - \frac{\sum_{m} \frac{N_0^2 16 L^2}{\pi^6 (2m-1)^4}}
{\left\{\sum_{m} \frac{N_0 4 L}{\pi^3 (2m-1)^2}\right\}^2} = 
\nonumber \\ 
&=& 1 - \frac{\frac{\pi^4}{96}}{\left\{\frac{\pi^2}{8}\right\}^2} = 
1 - \frac{2}{3} = \frac{1}{3} 
\;. \label {chaotoy3}
\end{eqnarray}

From the result in Eq.(\ref{chaotoy2}) we see that, 
for the particular density matrix we chose in 
Eq.(\ref{HN}), the intercept parameter, $\lambda({\bf K}=0)$, can 
still 
have the expected value for bosonic interferometry in the classical 
limit, 
i.e., $\lambda({\bf K}=0) = 0$, even for multi-modes, due to the 
combined 
effects of the density of states, $N_k$, and of the wave-functions, 
$\tilde{\psi}_{l}({\bf K}=0)$. However, in the second case, 
we see from Eq.(\ref{chaotoy3}), that 
$\lim_{L \rightarrow \infty} \lambda({\bf K}=0) = 1/3$. 
This is reflecting the relation between $L$ and $k$ coming from the 
shape 
of the wave function and boundary conditions in a finite system, 
as in Eq.(\ref{psi1}). Moreover, 
the difference among the two limits reflects the particular choice of 
the 
density matrix. In order to demonstrate this, we choose a much simpler 
example than in Eq.(\ref{HN}). For instance, let us choose $\rho_k = $ 
const., 
and proceed analogously as before. 
In this case, instead of Eq.(\ref{a+a2}) and (\ref{a+a+aa2}), we would 
have 
\begin{equation}
\langle a_{l}^{\dagger}a_{l}\rangle = 1 = N_l \; \;  ; \; \;  
\langle a_{l}^{\dagger} a_{l}^{\dagger} a_{l} a_{l}\rangle = 1
\; ,\label{2and4as}\end{equation}
As a consequence, we get for the sums in Eq.(\ref{sum2}), (\ref{sum4}) 
and 
Eq.(\ref{chaotoy3}) 

\begin{eqnarray}   
\left\{ \begin{array}{l}
            \mbox{ $\sum_{l}  |\tilde{\psi}_{l}(0)|^2 = \frac{4L}{\pi^3}$ } \; \\
	    \mbox{   } \; \\
	    \mbox{ $\sum_{l}  |\tilde{\psi}_{l}(0)|^4 = \frac{16 L^2}{\pi^6}$ 
}
             \end{array} \nonumber
             \right\} \Rightarrow \\
             %\Rightarrow 
	     \;\; \lambda({\bf K=0}) = 1 - 
	     \frac{\frac{16N_0^2L^2\pi^4}{96 \pi^6}}
	     {\left\{\frac{4N_0L\pi^2}{8 \pi^3}\right\}^2} = 
1 - \frac{2}{3} = \frac{1}{3}
\end{eqnarray}

From what we just saw above, we could say that, 
if we choose a different type of density matrix, i.e.,  
$\rho_k$ = const., for instance, the wave function does not seem to 
play 
any role, since we get $N_k$=const. 
Consequently, for this much simpler density matrix, 
we see that we get $\lambda \rightarrow 1/3$, independently on 
the energy of the states, $E_k$, 
and on the size of the 1-dimensional box, $L$.

\vfill
\newpage
\begin{center}
{\bf APPENDIX II}
\end{center}
\vskip1cm

For completeness, we will also derive below the so-called type-B 
$Q$-boson 
interferometry formulation, as defined in Ref.\cite{AGI}. For type-B 
$Q$-boson, 
the operators $b_i$ and $b_j$ satisfy the  
following commutation relations 
\begin{eqnarray}
&&b_{l}b_{l'}^{\dagger}-Q^{\delta_{l,l'}}
b_{l'}^{\dagger}b_{l}=\delta_{l,l'}Q^{-N_{l}}
\nonumber\\
&&b_{l}b_{l'}^{\dagger}-Q^{-\delta_{l,l'}}
b_{l'}^{\dagger}b_{l}=\delta_{l,l'}Q^{N_{l}}
\nonumber\\
&&[b_{l},b_{l'}]=[b_{l}^{\dagger},b_{l'}^{\dagger}]=0,
\nonumber\\
&&[\hat{N}_{l},b_{l'}]=-\delta_{l,l'}b_{l}
\nonumber\\
&&[\hat{N}_{l},b_{l'}^{\dagger}]=\delta_{l,l'}b_{l}^{\dagger},
\nonumber\\
&&[\hat{N}_{l},\hat{N}_{l'}]=0.
\label{commB}
\end{eqnarray}

In the above relations $Q$ is a parameter, which can be assumed within 
$[-1,1]$.
If we define it, following Ref.\cite{AGI}, as $Q = e^{i \theta}$ 
($0\le \theta < 2 \pi$) or, equivalently, 
$\cos(\theta)=\frac{1}{2}(Q+Q^{-1})$, then 

\begin{eqnarray}
&&\langle b_{l}^{\dagger}b_{l}\rangle=
\nonumber\\
&=&\frac{\exp(\frac{1}{T}(E_{l}-\mu))-1.}
{\exp(\frac{2}{T}(E_{l}-\mu))-2\cos(\theta)
\exp(\frac{1}{T}(E_{l}-\mu)+1} = 
\nonumber\\
&=& \frac{\exp[\frac{-1}{2T}(E_{l}-\mu)] 
\sinh[\frac{1}{2T}(E_{l}-\mu)]}
{\cosh[\frac{1}{T}(E_{l}-\mu)]-\cos(\theta)}
\label{bdagbB}
\end{eqnarray}
and 
\begin{eqnarray}
&&\langle b_{l}^{\dagger}
b_{l}^{\dagger}b_{l}
b_{l}\rangle=
\nonumber\\
&=&\frac{2 \cos(\theta)}
{\exp(\frac{2}{T}(E_{l}-\mu))-2\cos(2\theta)
\exp(\frac{1}{T}(E_{l}-\mu)+1}
\nonumber\\
&=&\frac{\cos(\theta) \exp[\frac{-1}{T}(E_{l}-\mu)]}
{\cosh[\frac{1}{T}(E_{l}-\mu)]-\cos(2\theta)}
\;.\label{bdagbdagbbB}
\end{eqnarray}

As before, the expectation value 
$ \langle \hat{b}^{\dagger}_{l}\hat{b}_{l'} \rangle $
is related to the occupation probability of the
single-particle state $l$,
$N^{(b)}_l$,  by a similar relation, i.e., 
\begin{equation}
\langle \hat{b}^{\dagger}_{l}\hat{b}_{l'} \rangle
=\delta_{l,l'}N^{(b)}_{l}
\; \;.\label{bdagabB2}\end{equation}

Then similar to the derivation of Eq.(16) and Eq.(20),we have 
\begin{equation}
P_1({\bf p})=\sum_{l} \langle b_{l}^{\dagger}b_{l}\rangle
\tilde{\psi}_{l}^*({\bf p})
\tilde{\psi}_{l}({\bf p})
\label{P1B}
\end{equation}
and 
\begin{eqnarray}
&&P_2({\bf p_1,p_2})=P_1({\bf p_1})P_1({\bf p_2})
+|\sum_{l}\tilde{\psi}_{l}^{*}({\bf p_1})
\tilde{\psi}_{l}({\bf p_2})|^2
\nonumber\\
&&+\sum_{l}
\tilde{\psi}_{l}^{*}({\bf p_1})
\tilde{\psi}_{l}^*({\bf p_2})
\tilde{\psi}_{l}({\bf p_1})
\tilde{\psi}_{l}({\bf p_2})
[\langle b_{l}^{\dagger}b_{l}^{\dagger}
b_{l}b_{l}\rangle -
2\langle b_{l}^{\dagger}b_{l}\rangle]
\nonumber\\
&&=
P_1({\bf p_1})P_1({\bf p_2})\!+\!\sum_{l,l'}\ N_{l} N_{l'} \!
\!\tilde{\psi}_{l}^*({\bf p_1}) \tilde{\psi}_{l'}^*({\bf p_2}) 
\tilde{\psi}_{l}({\bf p_2}) \tilde{\psi}_{l'}({\bf p_1})
\nonumber\\
&&\!
%\times
\!\left\{1-\delta_{l,l'}(1-\cos\theta)\!
\!\left[\frac{4 \cosh^2(\frac{E_l - \mu}{2T}) 
+ \frac{\cos\theta(\cos\theta - 1)}{\sinh^2(\frac{E_l - \mu}{2T})}} 
{\cosh(\frac{E_l - \mu}{T}) - \cos(2\theta)} \right] \right\} 
\!.\!\label{P2B}
\end{eqnarray}
Analogously to what was done at the end of Section II, we can also 
define 
a modified Wigner function for type-B $Q$-boson. 
The new  Wigner function for this case can be defined similarly as 
before, resulting in

\begin{eqnarray}
&&g^{(b)}({\bf x},{\bf K} ; {\bf y},{\bf K}) = g^{(b)} ({\bf x},{\bf 
K}) 
g^{(b)} ({\bf y},{\bf K}) 
- \nonumber\\
&&
\!\sum_{l}\! \left\{\!(N^{(b)}_{l})^2(1-\cos\theta)
\left[\!\frac{\!4 \cosh^2(\frac{E_l - \mu}{2T}) 
+ \frac{\cos\theta(\cos\theta - 1)}{\sinh^2(\frac{E_l - \mu}{2T})}\!} 
{\cosh(\frac{E_l - \mu}{T}) - \cos(2\theta)} \!\right]\!\!\right\}
\!.\!\!\label{wignerB}
\end{eqnarray}

Then the two-pion interferometry formula follows analogously 
to Eq.(\ref{c23rd}) for this case, i.e., 

\begin{equation}
C_2^{(b)} ({\bf p_1,p_2})= 1 +
\frac{\int \int e^{-i{\bf q}. ({\bf x}-{\bf y})}
g^{(b)} ({\bf x},{\bf K} ; {\bf y},{\bf K}) d{\bf x}d{\bf y}}
{\int g^{(b)} ({\bf x},{\bf p_1})d{\bf x} 
\int g^{(b)} ({\bf y},{\bf p_2}) d{\bf y}}
\end{equation}

\end {document}